\documentclass[conference]{IEEEtran}
\IEEEoverridecommandlockouts

\usepackage{cite}
\usepackage{amsmath,amssymb,amsfonts}
\usepackage{graphicx}
\usepackage{textcomp}
\usepackage{xcolor}
\usepackage{booktabs}
\usepackage{array}
\usepackage{url}

\begin{document}

\title{Market Attention and Information Quality in On-Chain Prediction Markets: Evidence from Polymarket}

\author{
\IEEEauthorblockN{
Davide Mancino\IEEEauthorrefmark{1}\IEEEauthorrefmark{2},
Hasret Ozan Sevim\IEEEauthorrefmark{2}\IEEEauthorrefmark{3},
Luca Pennella\IEEEauthorrefmark{4},
Huy Hoang Le\IEEEauthorrefmark{1}\IEEEauthorrefmark{2}
}

\IEEEauthorblockA{
\IEEEauthorrefmark{1}\textit{University of Milano-Bicocca}, Milan, Italy
\qquad
\IEEEauthorrefmark{2}\textit{University of Camerino}, Camerino, Italy\\
\IEEEauthorrefmark{3}\textit{Catholic University of the Sacred Heart}, Milan, Italy
\qquad
\IEEEauthorrefmark{4}\textit{University of Luxembourg}, Luxembourg\\
\texttt{davide.mancino@unimib.it \quad luca.pennella@uni.lu}\\
\texttt{\{hasretozan.sevim,huyhoang.le\}@unicam.it}
}
}

\maketitle

\begin{abstract}
This paper asks whether the leading on-chain prediction market, Polymarket, behaves more as an \emph{information market}, in which monetary stakes compress dispersed beliefs into a price that tracks the truth, or more as an \emph{attention market}, in which capital flows toward the most visible contracts rather than the most informative ones. Using trade-level data from Polymarket's central-limit-order-book (CLOB) markets from January 2024 through March 2026, and a one-sided exchange-contract convention that avoids double-counting maker and taker fills, we study the top 1,000 markets by taker-side dollar volume. For each, we build an \emph{attention} score from the within-sample percentile ranks of volume, transaction count, and active trader wallets, and an \emph{information-quality} score on resolved \emph{Yes/No} markets equal to one minus the absolute distance between the YES-token price seven days before resolution and the realised outcome. Placing markets on this attention/information plane answers the question in favour of the information-market reading: 68.2\% of analysis-sample volume sits in well-calibrated markets (average 7-day error of 0.03, falling to 0.02 one day before resolution), spread across 515 of 679 contracts. A narrower, identifiable cluster of 19 high-attention/low-information contracts captures 22.1\% of analysis-sample volume, but 79\% of that mass is the 2024 US Presidential Election, the single closest macro-political event of the window. Polymarket is therefore primarily an information aggregator; the attention component is real but small and concentrated in a handful of headline contracts whose outcomes are hardest to forecast ex ante.
\end{abstract}


\section{Introduction}
The case for prediction markets, from the price-as-aggregated-information argument to the modern literature on market scoring rules \cite{b1,b2}, rests on the premise that monetary stakes compress dispersed beliefs into a probability that is hard to beat. Survey-level evidence supports the claim that aggregate market prices outperform individual experts and structured polls \cite{b3}, and the platform-level calibration literature \cite{b4} typically reports that average errors are small and converge as resolution approaches. The recent on-chain wave extends prediction-market design to smart-contract-based venues, with Polymarket~\footnote{\url{https://polymarket.com}} and, to a lesser
extent, Kalshi multiplying the variety of contracts traded. Users post collateral to express probabilistic beliefs about elections and macro releases, and also to participate in news cycles or gain exposure to events with lottery-like payoff structures, so the platform-level reading naturally invites a finer-grained, segment-level test.

This paper asks a measurable question: \emph{is Polymarket more of an information market or more an attention market?} If the former, the platform behaves as the information-aggregation device the literature assumes. If the latter, prediction markets share important properties with sports betting or meme-asset trading: high engagement, high turnover, low informational content per dollar traded. The two readings are not exclusive, but at the platform level the data should adjudicate which is the dominant one.

We answer empirically. After documenting Polymarket's dominance within the on-chain segment, we restrict attention to the 1{,}000 deepest CLOB markets and decompose them along an attention dimension (dollar volume, transaction count, active trader wallets) and an information dimension (calibration error of the market price against the realised outcome). The data deliver a clear answer in favour of the information-market reading: well-calibrated markets account for 68.2\% of analysis-sample volume and 515 of 679 contracts (75.8\%), and the attention component is small (22.1\% of volume in 19 contracts) and heavily concentrated in the 2024 US election. Our contribution is a clean decomposition of where Polymarket activity sits in the attention/information plane, settling the binary question at the platform level and identifying the narrow contract set in which the attention reading remains relevant.

\section{Data and Why Polymarket}
\subsection{CLOB market structure}
Each Polymarket market is a central limit order book (CLOB) in which traders post orders against an ERC-1155 conditional-token pair (YES/NO), settled in USDC: a YES token pays one USDC on YES resolution and zero otherwise. Trades settle through the CTF Exchange smart contract, Polymarket's exchange contract for swapping Conditional Token Framework outcome tokens against collateral, which emits an \texttt{OrderFilled} event per fill. Two product variants coexist: \emph{standard} markets host a single binary contract, while \emph{negative-risk} markets bundle a set of mutually exclusive outcomes within one event (e.g.\ the candidate field of an election) and net YES tokens across outcomes so their prices sum to one, preventing the capital lock-up flat collateralisation would impose. Both share order-book mechanics and we pool them: the standard book contributes USD 16.27B of taker volume in 496{,}871 markets and the negative-risk book USD 12.14B in 126{,}156 markets.

\subsection{Data sources}
All trade-level data are reconstructed from on-chain primitives via Dune Analytics over January 1, 2024--March 31, 2026. We stop before Polymarket's April 28, 2026 CLOB V2 production cutover \cite{polymarketv2changelog}, preserving a single exchange architecture throughout. All code is publicly available on
our GitHub,\footnote{https://github.com/EconoDLT/PredictionMarketAnalysis} providing a reproducible implementation for testing, comparison, and
future extension.
We read the two V1 \texttt{OrderFilled} streams on Polygon: \texttt{ctfexchange\_evt\_orderfilled} (standard) and \texttt{negriskctfexchange\_evt\_orderfilled} (negative-risk). After taker-side filtering the tape contains 373.2M retained fills, 384.0M transactions, 2.39M active trader wallets and USD 28.41B of taker volume. Market metadata (questions, tags, resolution timestamps, outcomes) are joined from \texttt{polymarket\_polygon.market\_details}, and hourly mid-prices from \texttt{polymarket\_polygon.market\_prices\_hourly}, snapped to the closest hour within $\pm 24$h of $T-7$ days and $\pm 12$h of $T-1$ day.

\paragraph{Volume convention} The raw fill payload exposes both sides of each match, so naive sums double-count the economic exchange, a point flagged by Paradigm \cite{slivkoff2025paradigm} and corrected by the Dune ``Prediction Markets'' dashboard, the SOTA cross-venue reference \cite{dunepm2026}. We adopt the same one-sided convention: we keep the taker-side fill row, identified by the exchange contract in the event \texttt{taker} field, and record only the USDC leg of that row (\texttt{takerAmountFilled} if the taker is on the collateral side, \texttt{makerAmountFilled} otherwise). The \texttt{maker} field on the retained row identifies the initiating trader wallet, which we use for wallet-breadth counts. The convention is enforced identically in every paper-query, so all the figures here are produced from the same one-sided dollar leg.

\subsection{Why Polymarket}
Under this convention, we compute cross-venue volumes on Dune over January 1, 2024--March 31, 2026 using a single-sided taker-notional rule. For on-chain CLOB venues, including Polymarket, Opinion, Predict, and Limitless, we retain the taker-side \texttt{OrderFilled} event and count the collateral leg once; for Kalshi, we use \texttt{kalshi.trade\_report} and compute dollar notional as \texttt{$price \cdot contracts\_traded / 100$}; for Myriad, each trade is already a one-sided market action and we compute USD notional as \texttt{$value \cdot price / 10^{decimals}$}. This reconstruction reports USD 28.41B of Polymarket CLOB taker volume over the window, ahead of Kalshi (USD 24.41B), Opinion (USD 23.24B), Predict (USD 1.89B), Limitless (USD 1.42B), and Myriad (USD 0.39B). We centre the analysis on Polymarket for two reasons. First, the attention score~\eqref{eq:att} requires per-market wallet-breadth counts; Kalshi's \texttt{trade\_report} contains no user identifier, so the breadth dimension is mechanically unobservable, while Polymarket has 2.39M active wallets versus 247k for Opinion, 182k for Limitless and 65k for Predict, putting its observable retail base an order of magnitude ahead. Second, the information score~\eqref{eq:info} requires hourly mid-prices per market at $T{-}7$ days, which Dune publishes for Polymarket and not for Kalshi. 

\subsection{Volume concentration}
The contract universe is large but volume is extremely concentrated. The top three tags (Sports 39.48\%, Politics 29.01\%, Crypto 23.87\%) account for 92.4\% of dollar volume; the next twelve add only 6.0\%. The platform's economic mass sits in topics with broad public interest, motivating a segment-level rather than platform-average test.

\subsection{Defining attention}
For each of the top 1{,}000 markets by volume we compute three microstructure quantities: dollar volume $v_i$, distinct transactions $\tau_i$, and active trader wallets $w_i$. The attention score is the average within-sample percentile rank,
\begin{equation}
A_i = \tfrac{1}{3}\bigl[\mathrm{pr}(v_i) + \mathrm{pr}(\tau_i) + \mathrm{pr}(w_i)\bigr]\in[0,1].
\label{eq:att}
\end{equation}
Pre-selecting the universe before the percentile transformation anchors the threshold $\tau_A$ defined below to the analysis sample rather than to the full tape of $>10^5$ thinly-traded markets that would otherwise compress every score toward one. The three components capture complementary facets of this operational definition: $v_i$ the \emph{capital} a market attracts, $\tau_i$ the \emph{frequency} of interaction, and $w_i$ the \emph{breadth} of the user base. Combining all three penalises spurious one-dimensional spikes (a whale, a bot, an airdrop campaign). As an ex-post concentration check we compute the wallet-volume Herfindahl index, $H_i=\sum_j s_{ij}^2$, where $s_{ij}$ is wallet $j$'s share of market $i$ volume, together with the top-wallet share: in the high-attention cells the mean Herfindahl is 0.013--0.014 and the mean top-wallet share 5.8--6.0\%, indicating that high-attention markets reflect broad participation rather than whale concentration. We use percentile ranks rather than levels because the three quantities differ by orders of magnitude and are heavily right-skewed; and equal weights because no \emph{a priori} reason favours capital over breadth and the qualitative results survive any convex weighting we tried.

\subsection{Defining information quality}
The information score uses the realised outcome $y_i\in\{0,1\}$ and the YES-token price $p_i^{(h)}$ at horizon $h$ days before resolution:
\begin{equation}
Q_i^{(h)} = 1 - \bigl|p_i^{(h)} - y_i\bigr|\in[0,1].
\label{eq:info}
\end{equation}
The term $\bigl|p_i^{(h)}-y_i\bigr|$ is an absolute forecast error, consistent with prior prediction-market work that evaluates market accuracy by comparing market-implied forecasts with realised outcomes~\cite{BERG2008285}; we report its complement so that larger values denote higher information quality. Pipeline choices: (i) we snap each horizon to the closest hourly observation within $\pm 24$h of $T{-}7$ and $\pm 12$h of $T{-}1$, the tighter window reflecting higher trading intensity near resolution; (ii) prices are selected per \emph{(condition\_key, token\_id)} rather than per condition alone, so the score remains well-defined on negative-risk events with multiple YES tokens; (iii) markets whose metadata tags both YES and NO outcomes are nulled out. Of the 1{,}000 sample markets, 702 have a binary resolution within the window and 679 of those also have an in-tolerance $T{-}7$ price; the latter form the analysis sample and represent 81.3\% of top-1{,}000 dollar volume.

\subsection{Choice of the 7-day horizon}
The horizon $h$ is a central methodological choice. We use $h=7$ because prices near $T{-}1$ may reflect end-game arbitrage, position-squaring, and oracle-driven convergence rather than genuine forecasting. A seven-day snapshot is consistent with prior short-horizon prediction-market studies~\cite{BERG2008285} and preserves coverage of short-lived sports and weather markets that longer horizons would exclude. To guard against horizon-specific artefacts, we replicate every quadrant statistic at $h=1$ for all 679 markets.

\section{Results}
\subsection{The attention--information matrix}
\begin{figure}[tb]
\centerline{\includegraphics[width=0.9\columnwidth]{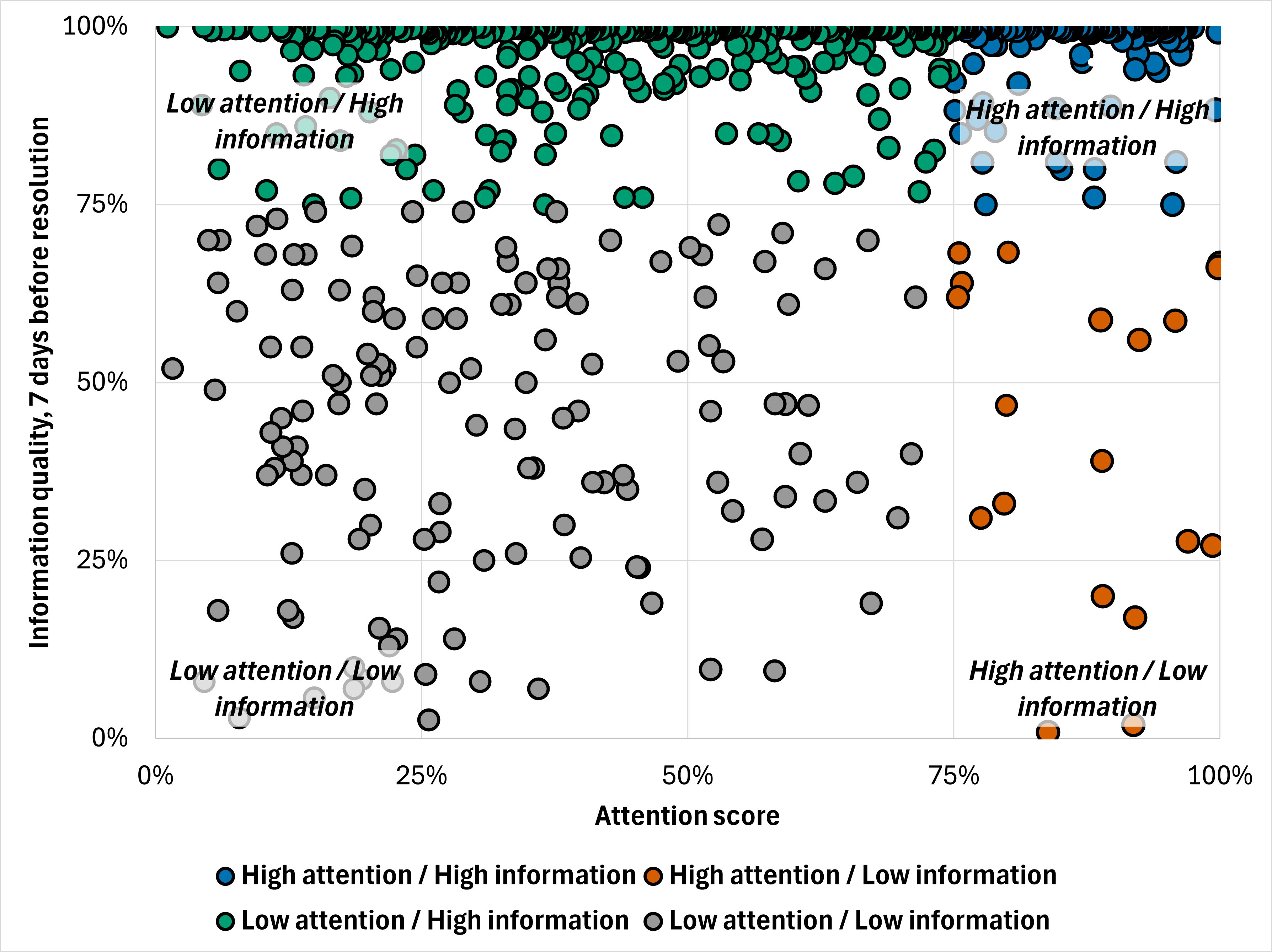}}
\caption{Attention score versus 7-day information quality for the 679 resolved binary markets in the analysis sample. Each point represents a market, coloured by attention--information quadrant.}
\label{fig:bubble}
\end{figure}

We classify each market as \emph{high attention} if $A_i\ge\tau_A$, with $\tau_A$ equal to the 75th within-sample percentile of the attention score ($\tau_A\approx 0.7556$), and as \emph{high information} if $Q_i^{(7)}\ge 0.75$. The thresholds define four quadrants whose volume share and accuracy are reported in Table~\ref{tab:matrix} and visualised in Fig.~\ref{fig:bubble}.

\begin{table*}[tb]
\caption{Attention--Information matrix for the top 1{,}000 Polymarket CLOB markets, by attention–information quadrant. 
%Volume shares are within the analysis sample of 679 resolved binary markets ($\Sigma v=$ USD 7.98B).
}
\label{tab:matrix}
\centering
\begin{tabular}{lrrrrrr}
\toprule
Quadrant & Markets & Volume (USD bn) & Vol.\ share & Avg.\ $Q_i^{(7)}$ & Avg.\ $|p^{(7)}-y|$ & Avg.\ $|p^{(1)}-y|$ \\
\midrule
High attention, high information & 152 & 3.28 & 41.2\% & 0.974 & 0.026 & 0.016 \\
High attention, low information  & 19  & 1.76 & 22.1\% & 0.452 & 0.548 & 0.293 \\
Low attention,  high information & 363 & 2.15 & 27.0\% & 0.967 & 0.033 & 0.017 \\
Low attention,  low information  & 145 & 0.78 & 9.8\%  & 0.455 & 0.545 & 0.322 \\
\bottomrule
\end{tabular}
\end{table*}

Fig.~\ref{fig:bubble} shows that markets are densely concentrated in the high-information region at both attention levels, whereas the high-attention/low-information quadrant contains only a small set of isolated contracts. Table~\ref{tab:matrix} quantifies this pattern. The two high-information rows together cover 515 of the 679 analysis-sample markets (75.8\% by count) and 68.2\% of dollar volume, with average 7-day errors of 0.026 and 0.033 falling to 0.016 and 0.017 by $T-1$: exactly the convergence pattern the prediction-market literature predicts. The high-attention/high-information cell alone, with 152 contracts and 41.2\% of volume, is sufficient to reject a strong attention-market reading: most of the platform's capital sits in well-trafficked markets that are also well-calibrated. A smaller high-attention/low-information cell coexists with it: 19 contracts capture 22.1\% of analysis-sample volume with an average 7-day error of 0.548, which partly converges to 0.293 by $T{-}1$. We examine what populates this cell next.

\subsection{What sits in the high-attention/low-information cell}
The cell is concentrated in a small, identifiable contract set. Two contracts alone, ``Will Donald Trump win the 2024 US Presidential Election?'' (USD 818.2M taker notional) and ``Will Kamala Harris win the 2024 US Presidential Election?'' (USD 437.6M), account for 71.3\% of the cell's volume; adding the popular-vote pair (Harris USD 80.4M, Trump USD 53.9M) brings the 2024-election YES tokens to 79.0\%. Seven days before the election the Trump YES traded at 0.667 against a realised 1, the Harris YES at 0.338 against a realised 0; the popular-vote contracts were inverted (Harris YES 0.729 against 0; Trump YES 0.277 against 1) and then converged within the last 24 hours (the Trump popular-vote contract had a T-1 price of 0.997). The remainder of the cell is dominated by ex-ante hard-to-forecast events: UEFA Champions League winner contracts (Inter Milan YES 0.412 against 0, USD 76.1M; PSG YES 0.587 against 1, USD 37.4M), Iran-related geopolitical contracts whose 7-day prices missed by 0.83--0.98 before snapping to the realised outcome within hours (US strikes Iran by Feb~28: YES 0.170 against 1; Khamenei out by Feb~28: YES 0.019 against 1), a TikTok-ban contract (YES 0.620 against 1), a Bitcoin-by-end-November contract (YES 0.690 against 0), and the Patriots/Chiefs/Seahawks Super Bowl contracts. These are precisely the contracts in which a one-week forecast horizon is genuinely difficult: a balanced presidential race, surprise geopolitical action, late-arriving political news, and high-variance sports finals. The cell does not look like an indictment of Polymarket as a forecasting venue; it looks like the contracts in which any forecaster would have struggled at $T{-}7$.

\subsection{Robustness}
Four features of the construction are worth flagging. First, anchoring at $T-7$ is conservative, in the sense that it deliberately strips out the mechanical end-game convergence; the symmetric $T-1$ exercise (Table~\ref{tab:matrix} last column) confirms calibration improves further once that channel is allowed in, with the calibrated cell mean falling to 0.016. Second, the attention threshold is the 75th within-sample percentile; lowering it expands the high-attention set with markets whose volume is small and increases the share of the high-attention/high-information cell. Third, the $Q\ge 0.75$ cutoff is permissive (it admits markets mispriced by up to 25 percentage points); tightening to $Q\ge 0.90$ leaves the relative ranking of the four quadrants unchanged. Fourth, the wallet-volume Herfindahl is 0.013--0.014 in the high-attention cells and 0.044--0.102 in the low-attention cells, so the high-attention rows reflect genuine retail breadth rather than whale capture.

A separate concern is selection in the other direction. We restrict to the 1{,}000 highest-volume markets for stable percentile ranks and informative price observations, and within them to the 679 with both a binary resolution and an in-tolerance $T{-}7$ price. This restriction conditions on the deepest contracts on the platform, which is where information aggregation has the best chance of operating; the 68.2\% well-calibrated share documented above is best read as a deep-market figure, not a platform-wide one.

\section{Discussion \& Conclusion}
The empirical pattern supports an information-market reading on most of Polymarket's dollar volume. 68.2\% of analysis-sample volume sits in well-calibrated markets, with average 7-day errors of 0.03 falling to 0.02 by $T{-}1$: the convergence pattern the literature attributes to prediction markets, delivered on the bulk of capital, including the macro-policy block (Fed rates, inauguration, NYC-mayoral) that populates the high-attention/high-information cell.

The 22.1\% high-attention/low-information mass is real but narrow, with three non-exclusive readings. First, hardest-to-forecast events: 79\% of the cell is the 2024 US Presidential Election, the closest US presidential race in recent history; the remainder is balanced football tournaments, surprise Iran action and fast-moving political news, so the residual error is largely irreducible at $T{-}7$. Second, an exposure-to-narrative component: the persistence of a 0.29 error into the last 24 hours suggests some demand for participation orthogonal to belief updating. Third, growth incentives (marketing, on-chain rewards) that draw retail flow toward contracts combining high salience with high outcome uncertainty. The first reading is consistent with the cell's narrow contract set and with the partial $T{-}1$ convergence.

Future work should test whether the attention--information decomposition generalises across on-chain venues and event categories, and examine how platform design and trader composition affect calibration.

%\section{Conclusion}
Using the pre-V2 Polymarket CLOB tape between January 2024 and March 2026, we classified the deepest 1,000 markets along attention and information dimensions. The platform displays the standard prediction-market pattern on most of its capital: 515 of 679 analysis-sample contracts (75.8\%) and 68.2\% of dollar volume are well-calibrated at $T{-}7$, with errors falling from about 0.03 to 0.02 by $T{-}1$. A narrower cluster of 19 high-attention/low-information contracts captures 22.1\% of volume and is concentrated, by 79\%, in the 2024 US Presidential Election and a handful of high-variance events. These findings are robust across forecast horizons, classification thresholds, and wallet concentration. Polymarket therefore functions primarily as an information aggregator while retaining an identifiable attention-driven component in headline contracts that are hardest to forecast. For researchers and platform designers, platform-level calibration masks meaningful contract-level heterogeneity, while engagement and informational output overlap substantially except in a small set of highly salient events.

\section*{Acknowledgment}
Luca Pennella's contribution was funded by the PayPal--FNR PEARL Chair in
Digital Financial Services, FNR grant reference 13342933/Gilbert Fridgen,
and by the FutureFinTech National Centre of Excellence in Research and
Innovation, FNR grant reference 16570468, with the support of Luxembourg's
Ministry of Finance.

\bibliographystyle{IEEEtran}
\bibliography{references}

@article{b1,
  author  = {Hanson, Robin},
  title   = {Combinatorial information market design},
  journal = {Information Systems Frontiers},
  volume  = {5},
  number  = {1},
  pages   = {107--119},
  year    = {2003},
  doi     = {10.1023/A:1022058209073}
}

@article{b2,
  author  = {Wolfers, Justin and Zitzewitz, Eric},
  title   = {Prediction markets},
  journal = {Journal of Economic Perspectives},
  volume  = {18},
  number  = {2},
  pages   = {107--126},
  year    = {2004},
  doi     = {10.1257/0895330041371321}
}

@article{b3,
  author  = {Arrow, Kenneth J. and Forsythe, Robert and Gorham, Michael
             and Hahn, Robert and Hanson, Robin and Ledyard, John O.
             and Levmore, Saul and Litan, Robert and Milgrom, Paul
             and Nelson, Forrest D. and Neumann, George R.
             and Ottaviani, Marco and Schelling, Thomas C.
             and Shiller, Robert J. and Smith, Vernon L.
             and Snowberg, Erik and Sunstein, Cass R. and Tetlock, Paul C.
             and Tetlock, Philip E. and Varian, Hal R. and Wolfers, Justin
             and Zitzewitz, Eric},
  title   = {The promise of prediction markets},
  journal = {Science},
  volume  = {320},
  number  = {5878},
  pages   = {877--878},
  year    = {2008},
  doi     = {10.1126/science.1157679}
}

@article{b4,
  author  = {Page, Lionel and Clemen, Robert T.},
  title   = {Do prediction markets produce well-calibrated probability forecasts?},
  journal = {The Economic Journal},
  volume  = {123},
  number  = {568},
  pages   = {491--513},
  year    = {2013},
  doi     = {10.1111/j.1468-0297.2012.02561.x}
}

@misc{slivkoff2025paradigm,
  author       = {Slivkoff, Storm},
  title        = {Polymarket volume is being double-counted},
  howpublished = {Paradigm Research},
  month        = dec,
  year         = {2025},
  note         = {Accessed: 2026-05-26},
  url          = {https://www.paradigm.xyz/2025/12/polymarket-volume-is-being-double-counted}
}

@misc{dunepm2026,
  author       = {{Dune Analytics}},
  title        = {{Prediction Markets dashboard---Taker Volume}},
  howpublished = {Dune query ID 7376904},
  year         = {2026},
  note         = {Accessed: 2026-05-26},
  url          = {https://dune.com/datadashboards/prediction-markets}
}

@misc{polymarketv2changelog,
  author       = {{Polymarket}},
  title        = {{Migrating to CLOB V2}},
  year         = {2026},
  howpublished = {\url{https://docs.polymarket.com/v2-migration}},
  note         = {Accessed: 2026-05-26}
}

@article{BERG2008285,
title = {Prediction market accuracy in the long run},
journal = {International Journal of Forecasting},
volume = {24},
number = {2},
pages = {285-300},
year = {2008},
note = {US Presidential Election Forecasting},
issn = {0169-2070},
doi = {https://doi.org/10.1016/j.ijforecast.2008.03.007},
url = {https://www.sciencedirect.com/science/article/pii/S0169207008000320},
author = {Joyce E. Berg and Forrest D. Nelson and Thomas A. Rietz}
}

\end{document}